\documentclass[10pt,twocolumn,journal,tse]{IEEEtran}
\usepackage{amsmath, mathtools, amssymb, amsthm, amsfonts}
\usepackage[numbers,comma,sort&compress]{natbib}
\usepackage{graphicx}
\usepackage{color}
\usepackage{caption}
\usepackage{subcaption}
\usepackage{float}
\usepackage{placeins}
\usepackage{listings}
\usepackage{multicol}
\usepackage{bicaption}
\usepackage{array}
\usepackage{cite}

\usepackage{lipsum}
\usepackage{graphicx}
\usepackage{url}
\usepackage{authblk}

\theoremstyle{definition}

\numberwithin{equation}{section}
\date{}
\title{Efficient Multichannel in XML Wireless Broadcast Stream}
\author[1]{ Arezoo  Khatibi*\thanks{Arezoo  Khatibi*, correspond email: \protect\url{arezoo.khatibi@grad.kashanu.ac.ir}, Faculty of Computer Science,  University of Kashan,  BLVD Ghotb Ravandi 6 kilometers, Kashan Iran.}}
\author[2]{Omid Khatibi}
\affil[1]{Faculty of Computer Science, University of Kashan, Kashan, Iran}
\affil[2]{Faculty of Mathematics, University of Vienna,Vienna, Austria} 

\begin{document}
\maketitle
\begin{abstract}
In this paper we recommend the use of multi-channel for XML data in wireless broadcasting. First we divide XML data into information units as bucket, then extract path information (XPath) for any unit and build an index tree from the data path. Finally, make wireless data stream with merging parts of index tree and parts of XML data in multichannel XML. Then, create a protocol that allows mobile users access to the wireless XML stream generated with our method. We study 11 channels in server side and 3 orthogonal channels in client side.
\end{abstract}

\section{Introduction}
\IEEEPARstart{A} sample of an XML stream is:
\begin{align*}
&<Root><a1><b1><c1/><c2/><c3/>\\
&</b1><b2><c4/><c5/><c6/></b2>\\
&<b3><c7/><c8/><c9/></b3></a1>\\
&<a2><b4><c10/><c11/><c12/></b4>\\
&<b5><c13/><c14/><c15/></b5><b6>\\
&<c16/><c17/><c18/></b6></a2><a3>\\
&<b7><c19/><c20/><c21/></b7><b8>\\
&<c22/><c23/><c24/></b8><b9><c25/>\\
&<c26/><c27/></b9></a3></root>
\end{align*}

And tree presentation from this XML stream is in the figure 1.
\begin{figure*}[htbp]
\centering
\includegraphics[width=1\linewidth]{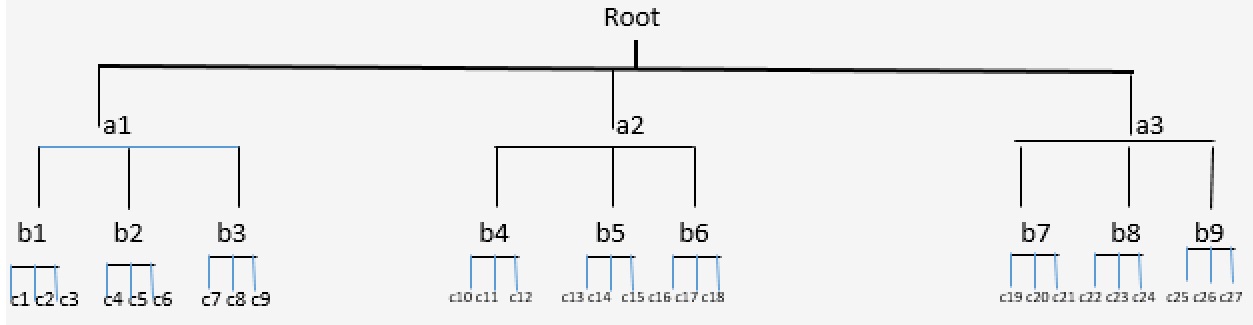}
        \caption{tree presentation from first sample of xml stream.}
        \label{Graph A}
\end{figure*}
\quad \quad \quad \quad \quad Yon Dohn Chung and Ji Leon Lee make a full study of an indexing for wireless broadcast XML data[1]. We develop it by using multichannel and multiradio. Data broadcast is widely used because of bandwidth limitations of wireless systems, where the server broadcasts data through a distribution channel and then the customers listen to the channel [3]. Customer can retrieve relevant information, furthermore, the client does not need to send a request to the server [4]. In order to retrieve information, we have an index broadcasting direct information of Published data items included minutes, seconds and milliseconds. Index in broadcast re-sults in energy saving. XML data is semi-structured and accessed with XPath (characteristic of path or XPath query). Previous methods of data broadcasting were adequate for flat data as database records accessed with attribute value of primary key and foreign key [2].
\section{RELATED WORK}
\subsection{(1, x) method}
(1, m) method satisfied flat data; however there is (1, x) in XML stream. In (1, x) method the index is represented as (XPath specification, start address, end address) and replicated x times during broadcast. But increased access time and wasted bandwidth. Therefore, we set partial and relevant index information in appropriate location in the broadcast stream. We divide data and index with using the control parameter replication into two parts: replicated and non-replicated.

\subsection{define tp strategy}

Tp stands for the tree of index and path of data. This building strategy of the replication stream,   replicates high level data path and high level index tree and put them before low level data path and low level index tree on the broadcast stream. It is suitable because of access time and tuning time for unique channel in compare of others method as pp, tt, (1, x).
\section{OUR PROPOSED METHOD}

We first implement, the XML tree traversal that we have noted in the introduction to the Java language along with the output of it.

\small
\begin{lstlisting}[language=Java]
1.	import java.io.File;
2.	import java.util.LinkedList;
3.	import javax.xml.parsers.DocumentBuilder;
4.	import javax.xml.parsers.DocumentBuilderFactory;
5.	import org.w3c.dom.Document;
6.	import org.w3c.dom.Element;
7.	import org.w3c.dom.Node;
8.	import org.w3c.dom.NodeList;
9.	/**
10.	*/
11.	public class Main {
12.	public static void main(String[] args) {
13.	try {
14.	DocumentBuilderFactory dbf = DocumentBuilderFactory.newInstance();
15.	DocumentBuilder db = dbf.newDocumentBuilder();
16.	File file = new File("C:\\New folder\\c.xml");
17.	if (file.exists()) {
•	Document doc = db.parse(file);
•	Node rootNode = doc.getFirstChild();
18.	//                System.out.println(rootNode.getNodeName());
19.	//                System.out.println(rootNode.getFirstChild().getNodeName());
20.	//                Sys-tem.out.println(rootNode.getFirstChild().getNextSibling().getNodeName());
21.	//                Sys-tem.out.println(rootNode.getFirstChild().getNextSibling().getNextSibling().getNodeName());
•	// node.getNodeName();   
•	// node.getFirstChild();    
•	// node.getNextSibling();  
•	System.out.println(getXML(rootNode));
•	LinkedList<Node> ll = getLL(rootNode);
•	for (int i=0;i<ll.size();i++){
•	Node myNode = ll.get(i);
•	System.out.println(myNode.getNodeName() + " index: " + ll.indexOf(myNode));
•	}
•	System.out.println();
•	// ll.get(index)   
•	// ll.indexOf(node)     
22.	}
23.	} catch (Exception e) {
24.	System.out.println(e);
25.	}
26.	}
27.	public static String getXML(Node root){
28.	if (root.getChildNodes().getLength() == 0)
29.	return "<"+root.getNodeName()+"/>";
30.	String a = "";
31.	a += "<"+root.getNodeName()+">";
32.	for (int i=0; i<root.getChildNodes().getLength();i++){
33.	a += getXML(root.getChildNodes().item(i));
34.	}
35.	a += "</"+root.getNodeName()+">";
36.	return a;
37.	}
38.	public static LinkedList<Node> getLL(Node root){
39.	LinkedList<Node> ll = new LinkedList<Node>();
40.	ll.add(root);
41.	Node thisNode = root.getFirstChild();
42.	while (!thisNode.equals(root)){
43.	ll.add(thisNode);
44.	if (thisNode.hasChildNodes()){
•	thisNode = thisNode.getFirstChild();
45.	} else if (thisNode.getNextSibling() != null){
•	thisNode = thisNode.getNextSibling();
46.	} else {
•	while (!thisNode.equals(root)
i.	&& thisNode.getNextSibling() == null)
•	thisNode = thisNode.getParentNode();
•	if (thisNode != root)
•	thisNode = thisNode.getNextSibling();
47.	}
48.	}
49.	return ll;
50.	}
51.	}
\end{lstlisting}
\begin{figure*}[htbp]
 \centering
      \includegraphics[width=1\linewidth]{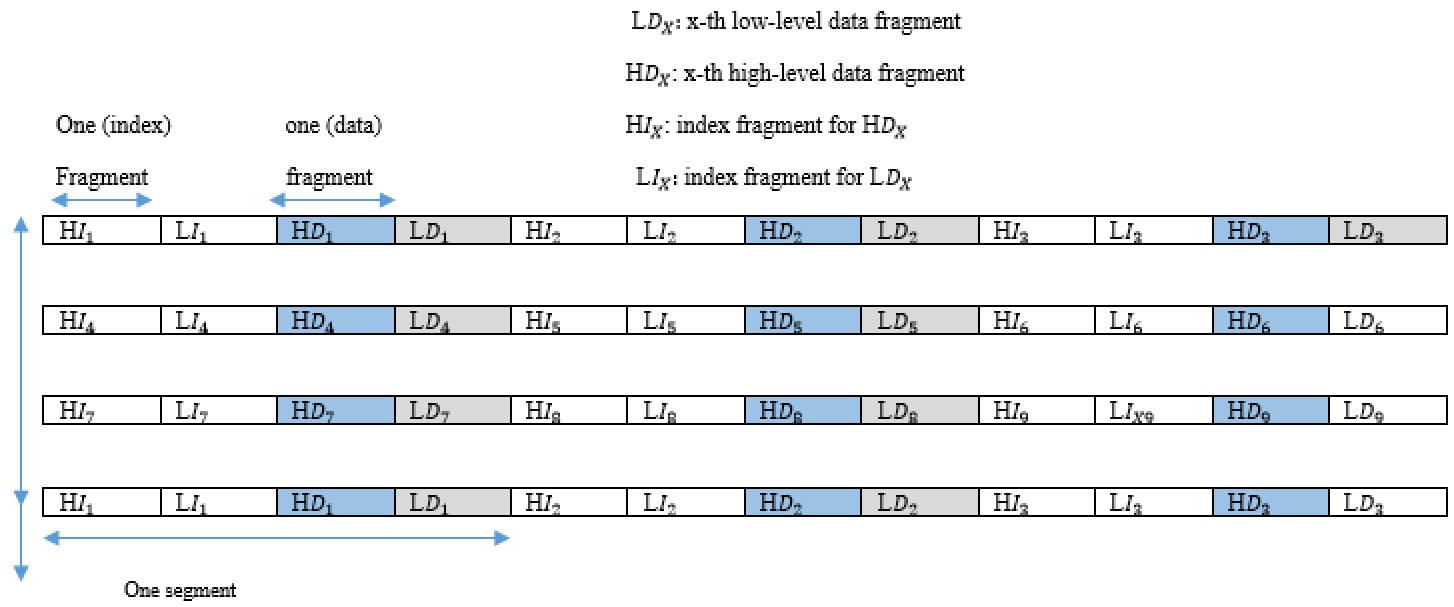}
        \centering
        \caption{Sequence d-link, c-link, s-link, h-link are data node address, the most left child node, the next sibling node and the relevant homolog node.}
        \label{Graph B}      
\end{figure*}
$<Root><a1><b1><c1/><c2/><c3/></b1><b2><c4/><c5/><c6/></b2><b3><c7/><c8/><c9/></b3></a1><a2><b4><c10/><c11/><c12/></b4><b5><c13/><c14/><c15/></b5><b6><c16/><c17/><c18/></b6></a2><a3><b7><c19/><c20/><c21/></b7><b8><c22/><c23/><c24/></b8><b9><c25/><c26/><c27/></b9></a3></Root>$

Root index: 0

a1 index: 1; b1 index: 2; c1 index: 3; c2 index: 4; c3 index: 5 ;b2 index: 6;  c4 index: 7; c5 index: 8  ; c6 index: 9 ; b3 index: 10; c7 index: 11;  c8 index: 12;  c9 index: 13 ; a2 index: 14 ; b4 index: 15; c10 index: 16 ; c11 index: 17;  c12 index: 18; b5 index: 19 ;c13 index: 20; c14 index: 21; c15 index: 22; b6 index: 23; c16 index: 24; c17 index: 25 ; c18 index: 26; a3 index: 27; b7 index: 28; c19 index: 29; c20 index: 30; c21 index: 31 ;b8 index: 32; c22 index: 33; c23 index: 34; c24 index: 35; b9 index: 36; c25 index: 37; c26 index: 38 ; c27 index: 39.

HI (high level index fragmentation): The sequence of index nodes in the stream that are lo-cated on the upper level of the index tree means. That sub trees existed with a root and h height related to top triangle of XML index but an XML index tree consists only one: HI. HI referred as the replicated index, in order to our all HI have same contents. 

LI (low level index fragment): The fraction of the wireless stream where index nodes stream exist between h+1 to H. We point LI as non-replication because content of the L1 to L9 are different. HD (high level data fragment): The sequence of data nodes in the stream located in the same path from root to the data node in h level. Following a string of (HI+LI+HD+LD) called as a segment. Wireless stream in the single channel is:
 \begin{table}[h!tbp]
  \centering
\begin{tabular}{|c|c|c|c|c|c|}
  \hline
      Ch1 & 1& 4 & 7& 1&4\\ \hline
    Ch2 & 2& 5 & 8& 2&5 \\ \hline
    Ch3 & 3& 6 & 9& 3&6\\ \hline
  \end{tabular}
  \caption{ Show the segments in each channel of multichannel. }
\end{table}
Instead we use one channel where we apply three channels .Put three segments 1, 4, 7 on channel 1 and put three segments 3, 6, 9 on channel 2 and we locate three segments 3, 6, 9 on channel 3 and as we know each segment consisted (HI+LI+HD+LD); then, in General, we can apply at most 3 orthogonal channels in client side and at most 11 channels in server side.

\section{PROTOCOL}
We use protocol 802.11 in client for receiving information. Each mobile customer can listen to only three channels at a moment.
\hspace*{-2cm}
\section{Analysis}
At first we want to utilize access time with our proposed method in TP strategy. In TP strategy HI replicate in form sub tree; therefore, average access time considered as follows:
Where n equals to the length of a high level in an XML tree in one channel .We consider a full tree with   H height and h referred as replication level and each node is in one bucket and do not consider download time after finding data. We assume value of index node in i level as linear function of f from i and k is a positive number , in our proposed method f(i)=k , we assume data node value in i level as g(i).
\hspace*{-2cm}
\subsection{For the average access time calculating by using TP strategy in multichannel case}
\hspace*{-2cm}
\begin{table}[h!tbp]
  \centering
\begin{tabular}{|c|c|}
  \hline
 Mark & Guide\\ \hline
$\Delta$LD &The total size of an LD\\ \hline
$\Delta$HD &The total size of an HD\\ \hline
$\Delta$LI &The total size of an LI\\ \hline
$\Delta$HI &The total size of an HI\\ \hline
$\Delta$INDEX &The size of total index in a broadcast sream\\ \hline
  $\Delta$DATA &The amount of data in a broadcast stream\\ \hline
f(i) &The size of an index node in a level i\\ \hline
g(i) &The size of the data node in a level i\\ \hline   
 X &The number of index replication in a (1,X) method\\ \hline 
 n &The fan out of a XML data and index trees\\ \hline
 H &The height of a XML data and index trees\\ \hline
 h &The replication level\\ \hline 
 w & Positive constant\\ \hline
 k & Positive constant\\ \hline 
  \end{tabular}
  \caption{ Show the symbols guide using in fourmula. }
\end{table}
\hspace*{-3cm}
\begin{equation}\label{eq:accesstime}
\begin{split}
AvgAT_{TP} =(\frac{1}{2}(\frac{n}{3})^h+1)k[\frac{(\frac{n}{3})^h-1}{\frac{n}{3}-1}+\frac{(\frac{n}{3})^{h-H}-1}{\frac{n}{3}-1}]
\\
+ \frac{1}{2}(\frac{n}{3})^h+1)w[h+\frac{(\frac{n}{3})^{h-H}-1}{\frac{n}{3}-1}],
\end{split}
\end{equation}
\begin{figure}[htbp]
      \includegraphics[width=\linewidth]{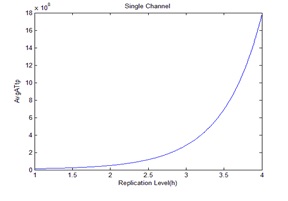}
        \caption{ In Shows access time of tp strategy in single channel.}
        \label{Graph C}
\end{figure}
\hspace*{-2cm}
\begin{table}[h!tbp]
  \centering
\begin{tabular}{|c|c|c|c|c|c|}
  \hline
n & h& H & K& w&$〖AvgAT〗_{TP}$\\ \hline
6 & 4& 7 & 3& 30&Single channel=
$107823\times〖10〗^9$\\ \hline
6 & 4& 7 & 3& 30&Multichannel=80784\\ \hline
  \end{tabular}
  \caption{ Get value of the access time used tp strategy with specified values in multichannel. }
\end{table}
\hspace*{-1cm}
\begin{figure}[htbp]
      \includegraphics[width=\linewidth]{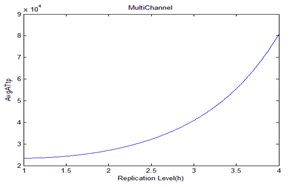}   
        \caption{ Shows access time of tp strategy in multichannel.}
        \label{Graph D}   
\end{figure}
\begin{figure}[htbp]
      \includegraphics[width=1\linewidth]{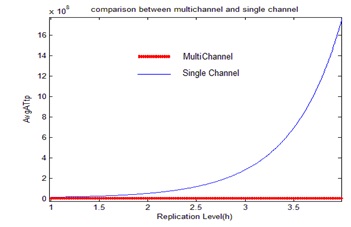}
        \caption{ Comparison access time in single channel and multichannel.}
        \label{Graph E}    
\end{figure}
\hspace*{-2cm}
\subsection{For  the index size calculating by using TP strategy in multichannel case}
\begin{equation}\label{eq:indexsize}
\begin{split}
〖\Delta{INDEX}〗_{TP}=(\frac{n}{3})^h(〖\Delta{HI}〗_{TP}+〖\Delta{LI}〗_{TP)})
\\
=k(\frac{n}{3})^h[\frac{(\frac{n}{3})^{h}-1}{\frac{n}{3}-1}+
\frac{(\frac{n}{3})^{h-H}-1}{\frac{n}{3}-1}]
\end{split}
\end{equation}
\hspace*{-1cm}
\begin{table}[h!tbp]
  \centering
\begin{tabular}{|c|c|c|c|c|}
  \hline
n & h& H & K& $\Delta{INDEX}_{TP}$\\ \hline
6 & 4& 7 & 3& Single channel=
1174176\\ \hline
    6 & 4& 7 & 3& Multichannel=1056\\ \hline
  \end{tabular}
  \caption{ get value of the index used tp strategy with specified values in single channel and  multichannel.}
\end{table}
\hspace*{-1cm}
\begin{figure}[htbp]
      \includegraphics[width=1\linewidth]{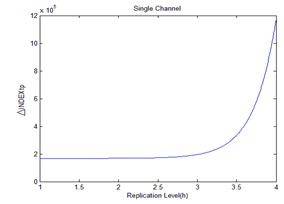}
        \caption{ Shows of index size with tp strategy in single channel.}
        \label{Graph F}    
\end{figure}
\hspace*{-2cm}
\begin{figure}[htbp]
      \includegraphics[width=1\linewidth]{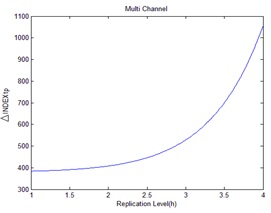}
        \caption{Shows of index size with  tp strategy in multichannel.}
        \label{Graph G}     
\end{figure}
\hspace*{-2cm}
\begin{figure}[htbp]
      \includegraphics[width=1\linewidth]{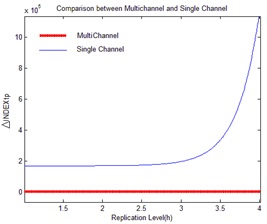}
        \caption{Comparison index size of tp strategy between single channel and multichannel.}
        \label{Graph H}
\end{figure}
\hspace*{-2cm}
\subsection{For the average tuning time calculating by using TP strategy in multichannel case }
\hspace*{-2cm}
\begin{equation}\label{eq:tuningtimemultichanel}
\begin{split}
〖AvgTT〗_{TP}=1+〖\Delta{HI}〗_TP+〖\Delta{LI}〗_TP
\\
=1+k[\frac{(\frac{n}{3})^{k}-1}{\frac{n}{3}-1}+
\frac{(\frac{n}{3})^{h-H}-1}{\frac{n}{3}-1}]
\end{split}
\end{equation}
\hspace*{-2cm}
\begin{table}[h!tbp]
  \centering
\begin{tabular}{|c|c|c|c|c|}
  \hline
n & h& H & K& $〖AvgTT〗_{TP}$\\ \hline
6 & 4& 7 & 3& Single channel=
907\\ \hline
    6 & 4& 7 & 3& Multichannel=67\\ \hline
  \end{tabular}
  \caption{ get value of the tuning time used tp strategy with specified values in multichannel.}
\end{table}
\hspace*{-2cm}
\begin{figure}[htbp]
      \includegraphics[width=1\linewidth]{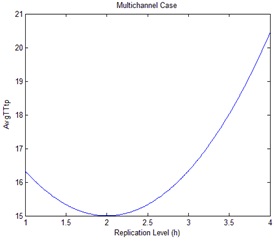}
        \caption{tuning time in multichannel case.}
        \label{Graph I}
\end{figure}
\hspace*{-2cm}
\begin{figure}[htbp]
      \includegraphics[width=1\linewidth]{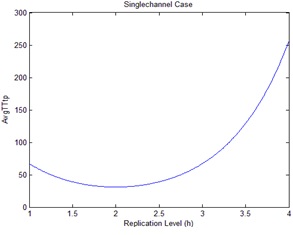}
        \caption{tuning time in single channel case.}
        \label{Graph J}      
\end{figure}
\begin{figure}[htbp]
      \includegraphics[width=\linewidth]{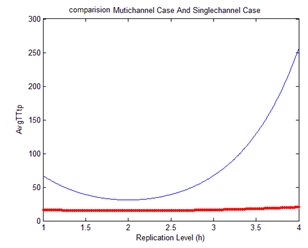}   
        \caption{ comparison tuning time between multichannel and single channel.}
        \label{Graph K}      
\end{figure}
\subsection{For the data size calculating by using TP strategy in multichannel case}
\hspace*{-2cm}
\begin{equation}\label{eq:datasize}
\begin{split}
〖\Delta{DATA}〗_{TP}
=(\frac{n}{3})^h(〖\Delta{HD}〗_{TP}+〖\Delta{LD}〗_{TP} )
\\
=(\frac{n}{3})^k[\sum_{i=1}^h g(i)+\sum_{i=1}^{h-H} (g(i)\times (\frac{n}{3})^{i-1}]
\end{split}
\end{equation}
\begin{equation}\label{eq:datasize2}
\begin{split}
=w (\frac{n}{3})^k[h+\frac{(\frac{n}{3})^{h-H}-1}{(\frac{n}{3})-1}]
\end{split}
\end{equation}
\hspace*{-2cm}
\begin{table}[h!tbp]
  \centering
\begin{tabular}{|c|c|c|c|c|}
  \hline
      n & h& H & w& $〖\Delta{DATA}〗_{TP}$\\ \hline
    6 & 4& 7 & 30& Single channel=
1827360\\ \hline
    6 & 4& 7 & 30& Multichannel=5280\\ \hline
  \end{tabular}
  \caption{ get value of the data size used tp strategy with specified values in multichannel.}
\end{table}
\begin{figure}[htbp]
      \includegraphics[width=\linewidth]{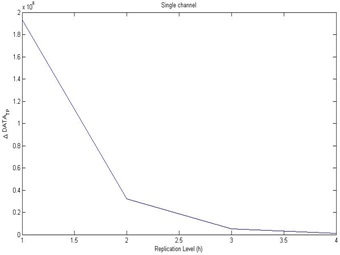}
        \caption{ data size of tp strategy in single channel.}
        \label{Graph L}
\end{figure}
\hspace*{-2cm}       
\begin{figure}[htbp]
      \includegraphics[width=\linewidth]{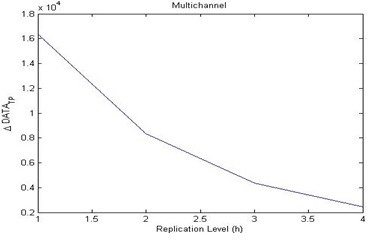}
        \caption{ data size of tp strategy in multichannel.}
        \label{Graph M}          
\end{figure}
\hspace*{-2cm}
\begin{figure}[htbp]
      \includegraphics[width=1\linewidth]{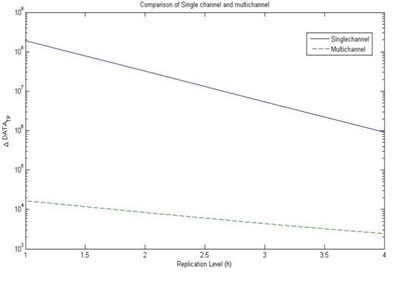} 
        \caption{compare data size of tp strategy between single channel and multichannel.}
        \label{Graph N}    
\end{figure}
\hspace*{-2cm}
\section{Conclusion}
With increasing number of channels, data is broadcast faster in air. In this model, we must have a multiradio or have single device with several radio. We observed that tuning time reduced with increasing number of channel or reducing number of nodes. In index size with increasing channel or decreasing number of nodes, index size increase faster and displayed more mutant. While accessing time in multichannel case with increasing replication level we see access time decreases faster. In future work we recommend a method to get rid of trav-erse tree structure in multichannel xml wireless broadcast.
Nonfinancial and no fund.

\begin{IEEEbiographynophoto}{Arezoo Khatibi}

 correspond email: \protect\url{arezoo.khatibi@grad.kashanu.ac.ir}, Faculty of Computer Science,  University of Kashan, BLVD Ghotb Ravandi 6 kilometers, Kashan, Iran. Her research interests cover most spectral Internet of Thing, distributed system, parallel programming. Khatibi attended the Nemo Internet of Thing at University of Vienna (2016).  
 [{\includegraphics[width=1in,height=1.25in,clip,keepaspectratio]{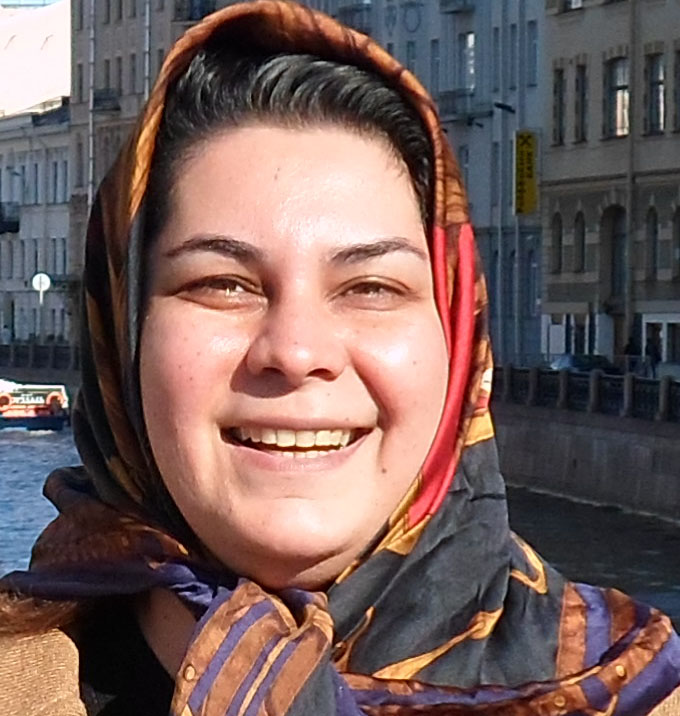}}]{Arezoo Khatibi}

\end{IEEEbiographynophoto}
\end{document}